\documentclass[12pt]{article}
\usepackage[dvipdfm]{graphicx}
\usepackage{amssymb}
\usepackage{amsmath}
\usepackage{eufrak}
\newcommand{\bi}{\bibitem}
\newcommand{\bea}{\begin{equation}}
\newcommand{\eea}{\end{equation}}
\newcommand{\be}{\begin{eqnarray}}
\newcommand{\ee}{\end{eqnarray}}

\def\hbar#1{\backslash\hspace{-2mm}#1}

\catcode`\@=11
\def\lsim{\mathrel{\mathpalette\@versim<}}
\def\gsim{\mathrel{\mathpalette\@versim>}}
\def\@versim#1#2{\vcenter{\offinterlineskip
\ialign{$\m@th#1\hfil##\hfil$\crcr#2\crcr\sim\crcr } }}
\catcode`\@=12

\def\2tvec#1#2{
\left(
\begin{array}{c}
#1  \\
#2  \\
\end{array}
\right)}

\def\mat2#1#2#3#4{
\left(
\begin{array}{cc}
#1 & #2 \\
#3 & #4 \\
\end{array}
\right) }

\def\Mat3#1#2#3#4#5#6#7#8#9{
\left(
\begin{array}{ccc}
#1 & #2 & #3 \\
#4 & #5 & #6 \\
#7 & #8 & #9 \\
\end{array}
\right) }

\def\3tvec#1#2#3{
\left(
\begin{array}{c}
#1  \\
#2  \\
#3  \\
\end{array}
\right)}

\def\hbar#1{\backslash\hspace{-2mm}#1}

\numberwithin{equation}{section}

\begin{document}

\begin{titlepage}
\begin{flushright}
KANAZAWA-11-16, P11051
\end{flushright}

\begin{center}

\vspace{1cm}
{\large\bf A light Scalar Dark Matter for CoGeNT and DAMA\\
in $D_6$ Flavor Symmetric Model }
\vspace{1cm}

Yuji Kajiyama$^{a}$\footnote{kajiyama-yuuji@akita-pref.ed.jp}
,
Hiroshi Okada$^{b,c}$\footnote{HOkada@Bue.edu.eg}
 and
Takashi Toma$^{d,e}$\footnote{t-toma@hep.s.kanazawa-u.ac.jp}
\vspace{5mm}

{\it%
$^{a}${Akita Highschool, Tegata-Nakadai 1, Akita, 010-0851, Japan}\\
$^{b}${Centre for Theoretical Physics, The British University in 
Egypt,\\[-1.5mm] El Sherouk City, Postal No, 11837, P.O. Box 43, Egypt}\\
$^{c}${School of Physics, KIAS, Seoul 130-722, Korea}\\
$^{d}${ Institute for Theoretical Physics, Kanazawa University, Kanazawa, 920-1192, Japan}\\
  $^{e}${Max-Planck-Institut f\"ur Kernphysik, Postfach 103980, 69029 Heidelberg, Germany}}
  
  \vspace{8mm}

\abstract{
We try to interpret a very light dark matter with mass of $
5\sim10\ {\rm GeV}$ which is in favor of the recent experiments reported
 by CoGeNT and DAMA, in a non-supersymmetric extension of
 radiative seesaw model with a family symmetry $D_6 \times
 \hat{Z_2} \times Z_2$. We show that a $D_6$ singlet real scalar field can be a promising dark matter candidate, and it gives the elastic cross section $\sigma\simeq 
7\times10^{-41}\ {\rm cm^2}$ which is required by these experiments. 
Our dark matter interacts with a $D_6$ singlet scalar Higgs boson, which couples only to quark sector.
The dark matter-nucleon cross section and new decay mode $h\to$DM DM can be large if the 
standard model Higgs boson $h$ is light. 
The Higgs phenomenology is also discussed.}

\end{center}
\end{titlepage}

\setcounter{footnote}{0}

\section{Introduction}

The existence of the dark matter (DM) in the Universe has been established 
by measurements. 
The WMAP experiment tells us that the amount of the DM is considered about 
23$\%$ of energy density of the Universe \cite{wmap}, while the baryon density is about 4 $\%$. 
Recently, it is reported that DM mass is favored in the range $5-10\ {\rm GeV}$ by the direct detection experiments of Contact Germanium Detector (CoGeNT) \cite{cogent} and DArk MAtter (DAMA) \cite{Bernabei:2008yi}.

If an asymmetry between DM and anti-DM is correlated to baryon
asymmetry, the above measurements suggest that DM is about 5-6 times heavier than baryon
(nucleon) due to the ratio of the two asymmetries.  
Since it is in fact a natural way to interpret that DM mass should be $5-10\ {\rm GeV}$, 
many authors have been working with many scenarios of this direction, which is called ``Asymmetric Dark Matter" scenarios \cite{models}.

In this letter, instead, we try to interpret the lightness of DM in a non-supersymmetric extension of
 radiative seesaw model with a family symmetry based on $D_6 \times
 \hat{Z_2} \times Z_2$ (see also a review of non-abelian discrete symmetry \cite{review}). There are many works based on $D_6$ flavor
 symmetry \cite{Dihedral} and radiative seesaw mechanism
 \cite{radiative-seesaw}.  The relation between DM and flavor symmetry
 is discussed in Ref.\cite{Hirsch:2010ru, Esteves:2010sh, Meloni:2010sk,
 Boucenna:2011tj}. We show that a $D_6$ singlet real scalar field could be a promising DM candidate,
and gives the elastic cross section $\sigma\simeq 7\times10^{-41}\ {\rm cm^2}$ which is required by these experiments.
Our DM interacts with nucleons through the t-channel diagram mediated by $D_6$ singlet scalar Higgs boson $\phi_S$, 
and the resulting elastic cross section can be large for relatively light Higgs boson. At the same time, 
the SM Higgs boson decays into two DM particles. The branching ratio of the new decay mode becomes large 
for the case of light Higgs boson. Therefore, the light SM Higgs boson $(m_h \sim 120 \rm{GeV})$ is 
favored in our model. 

Recently, ATLAS \cite{atlas} and CMS \cite{cms} reported an upper bound of the standard model (SM),
in which the Higgs mass is excluded in the range between $145$ GeV and $466$ GeV.   
However our SM Higgs is relaxed due to the mixing in the multi-Higgs sector. 
We show that the mass of our SM Higgs comes into the allowed range in our parameter space. And also we show our branching ratios, in which the new mode that SM Higgs decays into two light DM particles is depicted.

This paper is organized as follows. In section 2, we review our model 
briefly. In section 3 and 4, we analyze the Higgs potential and the DM mass 
from WMAP, respectively. In section 5, We discuss the direct detection of DM for CoGeNT/DAMA. In section 6, we analyze the Higgs phenomenologies and discuss our SM Higgs.
Section 7 is devoted to conclusions.

\section{The Model}
We consider a non-supersymmetric extension
of radiative seesaw model with a family symmetry based on $D_6\times
\hat{Z}_2\times Z_2$ \cite{Kajiyama:2006ww}. 
We introduce three Higgs doublets $\phi_{I,S}$, three inert 
doublets $\eta_{I,S}$, and one inert singlet $\varphi$, where 
$I=1,2$ and $S$ denote $D_6$ doublet and singlet, respectively. We assign 
charges of $SU(2)_L\times U(1)_Y$ and $D_6\times \hat{Z}_2\times Z_2$ to each field in specific way shown in 
Table \ref{a1}. 
\begin{table}[thb]
\begin{center}
\scalebox{0.9}[0.9]{
\begin{tabular}{|c|cccccc|cccc|c|} \hline
 & $L_S$    & $n_S$    & $e^c_S $ & $L_I$&$n_I$&$e^c_I$
 & $\phi_S$ & $\phi_I$ & $\eta_S$ & $\eta_I$ & $\varphi$ \\ \hline
 $\!\!SU(2)_L\times U(1)_Y\!\!$ 
 & $\!\!({\bf 2}, -1/2)\!\!$  &  $\!\!({\bf 1}, 0)\!\!$  &  $\!\!({\bf 1}, 1)\!\!$
 & $\!\!({\bf 2}, -1/2)\!\!$&  $\!\!({\bf 1}, 0)\!\!$
 &  $\!\!({\bf 1}, 1)$ & $\!\!({\bf 1},-1/2)\!\!$ & $\!\!({\bf
 1},-1/2)\!\!$ & $\!\!({\bf 1},-1/2)\!\!$ & $\!\!({\bf 1},-1/2)\!\!$ &
 $\!\!({\bf 1}, 0)\!\!$
  \\ \hline
 $D_6$ & ${\bf 1}$  &  ${\bf 1}'''$  &  ${\bf 1}$
 & ${\bf 2}'$&  ${\bf 2}'$&  ${\bf 2}'$ & ${\bf 1}$ & ${\bf 2}'$
 & ${\bf 1}'''$ & ${\bf 2}'$ & {\bf 1}
 \\ \hline
 $\hat{Z}_2$
 & $+$ &$+$  & $-$  & $+$ 
 &$+$  & $-$ & $+$ & $-$ & $+$ & $+$ & $+$
  \\ \hline
   $Z_2$
 & $+$ &$-$  & $+$  & $+$ &$-$  & $+$ 
 & $+$ & $+$ & $-$ & $-$ & $-$
   \\ \hline
\end{tabular}
}
\caption{The $SU(2)_L\times U(1)_Y\times D_6 \times \hat{Z}_2\times Z_2$ 
assignment for the leptons and the Higgs bosons.  The subscript $S$ 
indicates a $D_6$ singlet and the subscript $I$ which run from $1$ to $2$ indicates 
a $D_6$ doublet. $ L_I$ and $L_S$ denote the $SU(2)_L$-doublet 
leptons,
while $e^c_I$, $e^c_S$, $n_I$ and $n_S$ are the $SU(2)_L$-singlet leptons.
}
  \label{a1}
\end{center}
\end{table}
All quarks and a Higgs
doublet $\phi_S$ are assigned to be full singlet under the family
symmetry $D_6\times \hat{Z}_2\times Z_2$. Thus the 
quark sector is basically the same as the SM, and $\phi_S$ is regarded
as 
the SM Higgs
in the quark sector. No other Higgs bosons can couple to the quark sector 
at tree-level and then tree-level flavor changing neutral currents 
(FCNCs) do not exist in the quark sector.
The right-handed neutrinos $n_I, n_S$, 
the inert Higgs doublets $\eta_I, \eta_S$ and $\varphi$ are odd under
the $Z_2$ symmetry which plays the role of $R$-parity in supersymmetric models.
Although the field contents of our model are same as \cite{Kajiyama:2011fe}, 
the $Z_2$ charge of $\varphi$ is different. 
This ensures the stability of of $\varphi$, which is our DM candidate.  
As shall be discussed later, the gauge singlet $\varphi$ is found to be a good DM candidate, which plays 
an important role to explain the direct detection measurements of DM reported by CoGeNT
and DAMA. 

The most general renormalizable $D_6 \times \hat{Z}_2 \times 
Z_2$ invariant 
Yukawa interactions in the lepton sector 
are found to be
\bea
{\cal L}_{Y} =\sum_{a,b,d=1,2,S}~\left[
Y_{ab}^{ed} (L_{a} i\sigma_2\phi_d) e^c_{b} 
+Y_{ab}^{\nu d} (\eta_d^\dag L_{a}) n_{b} \right]
- \sum_{I=1,2}\frac{M_1}{2}n_{I}n_{I}-
 \frac{M_S}{2}n_{S}n_{S}.
 \label{wL}
\eea
We assume that the electroweak symmetry breaking is caused by the vacuum expectation values (VEVs) 
$\langle \phi_1 
\rangle=\langle \phi_2 \rangle\equiv
v_D/2~,\langle \phi_S\rangle=v_S/\sqrt{2},~V^2\equiv v_D^2+v_S^2=(246~{\rm 
GeV})^2$ and $\langle \eta_{I,S}\rangle=\langle \varphi\rangle=0$
\cite{okada1}. 
The form of the mass matrix of charged leptons is determined by the flavor symmetry and 
VEV alignments. See Ref.(\cite{Kajiyama:2006ww}) for detalis. 
In the neutrino sector, Yukawa couplings in the mass eigenstates
are given by
\be
Y^S&=&U_{eL}^TY^{\nu S},~
Y^\pm = \frac{1}{\sqrt{2}}U_{eL}^T(Y^{\nu 1}\pm Y^{\nu 2}),\\
Y^S& \simeq &
\left(\begin{array}{ccc}
\!\! 0 \!  &\!  0\!  & \!\!h_{3}\!\! \\
\!\! 0 \!  &\!  0\!  &  \sqrt{2}\epsilon_e h_{3}\!\!  \\
\!\! 0 \!  &\!  0\!  &\!\! 0 \!\! 
\end{array}\right),~\label{yukL}
Y^+ \simeq 
\left(\begin{array}{ccc}
\!\! \frac{h_{4}-2\epsilon_e h_2}{\sqrt{2}}  & 
\!\! \frac{h_{4}}{\sqrt{2}} &  0  \\
\!\! h_2  + \epsilon_e h_{4}  & 
\!\!  \epsilon_e  h_{4}  &  0  \\
\!\! 0  & \!\! h_2  &  0 
\end{array}\right)
\label{yukH},~
Y^- \simeq
\left(\begin{array}{ccc}
\frac{h_{4}}{\sqrt{2}} &
\frac{-h_{4}-2\epsilon_e h_2}{\sqrt{2}}
 &0 \\
\!\!\epsilon_e h_{4} \!\!& 
\!\!h_2  - \epsilon_e h_{4}\!\! &0 \\
\!\!-h_2 \!\!& \!\!0\!\! & 0
\end{array}\right)
\label{yukm}, 
\ee
where the Dirac Yukawa couplings $h_i~(i=2,3,4)$ are of order one,
$\epsilon_e\equiv m_e/(\sqrt{2}m_\mu)$ and $U_{eL}$ is diagonalization
matrix for the mass matrix of charged lepton. Notice that the $D_6$ singlet
right-handed neutrino $n_S$ couples only with $L_S$ and $\eta_S$.
In the present model Dirac neutrino mass term does not exist because of the
exact $Z_2$ symmetry and
vanishing VEVs of $\eta_{I,S}$. Thus, although canonical seesaw mechanism does not
work for generating light Majorana neutrino masses, radiative seesaw mechanism 
works at one-loop level\cite{radseesaw}. 
In this mechanism, Majorana mass is 
proportional to $h_i^2 \kappa V^2 M/(16 \pi^2 (M^2-m_{\eta}^2))$, where 
$M$ is heavy Majorana neutrino mass ($M_1$ or $M_S$) and $\kappa$ denotes typical coupling constant of 
non self-adjoint terms such as $(\phi^\dag \eta)^2$ in the Higgs
potential. 
Since a new $U(1)$ symmetry appears in the limit of $\kappa \to 0$, 
it is natural to suppose that the small breaking of the $U(1)$ symmetry ensures the 
smallness of neutrino masses. 
Therefore, we take $\kappa \ll 1$, $M_{1,S}={\cal O}({\rm TeV})$ and 
$h_i={\cal O}(1)$ to give neutrino masses. 

\section{Higgs Potential } 
In this section, we analyze the Higgs potential in our model. 
As discussed in Refs.\cite{Kajiyama:2006ww,okada1}, the Higgs potential 
consists of $D_6$ symmetric and breaking terms. 
Since the $D_6$ invariant Higgs potential has an accidental global $O(2)$ 
symmetry, 
the breaking terms must be introduced in order to forbid massless Nambu-Goldstone 
(NG) bosons. 
Essentially, such soft $D_6$ breaking terms are mass terms of the Higgs 
bosons.  
For the potential of $(\phi_I,\phi_S)$, the soft $D_6$  
breaking mass terms \cite{okada1} are given by
\bea
V(\phi)_{soft}=
\mu^2_2(\phi^{\dagger}_2\phi_1+\phi^{\dagger}_1\phi_2)
+\left(\mu^2_4\phi^{\dagger}_S(\phi_1+\phi_2)+h.c.\right), 
\label{softD6break}
\eea
where $\mu^2_2$ is real while  $\mu^2_4$ is complex in general. 
The mass term of $(\phi_I,\phi_S)$ is dominated by Eq.(\ref{softD6break}), 
and 
subdominantly given by $D_6$ invariant terms of order $V^2$. 
One finds that the $D_6$ breaking terms Eq.(\ref{softD6break}) preserve 
the minimal symmetry $S_2$ under $\phi_1\leftrightarrow\phi_2$.
The key point is that the $S_2$ invariance is required not only to ensure 
the vacuum alignment $\langle \phi_1 \rangle=\langle \phi_2 \rangle \neq 0$ 
but also to forbid NG bosons which violate the electroweak precision test 
of the SM.

Since the Higgs potential of $\phi_{I,S}$ and $\eta_{I,S}$ are analyzed in 
Ref.\cite{Kajiyama:2006ww}, we do not explicitly show that here again. 
In the present model, the new field $\varphi$ is introduced and 
it plays an important role in our analysis. Therefore we explicitly show 
the potential including $\varphi$. 
The most general renormalizable $D_6 \times \hat{Z}_2 \times 
Z_2$ invariant Higgs potential of $\varphi$ is given by
\be
V(\varphi)&=&m^2_2\varphi^2+\lambda_1\varphi^4,\\
V(\phi,
\varphi)&=&
\lambda_2(\phi^{\dagger}_S\phi_S)\varphi^2+\lambda_3(\phi^{\dagger}_I\phi_I)\varphi^2,\label{higgs-int}\\
V(\eta, \varphi)&=&V(\phi, \varphi)(\phi\rightarrow \eta),
\ee
where all parameters are considered to be real without loss of generality.
By using the decomposition of $SU(2)_L$ doublets $\phi_{I,S}$, 
\be
\phi_I=\frac{1}{\sqrt2}
\left(\begin{array}{c}
v_D/\sqrt2+\rho_I+i\sigma_I\\
\sqrt2\phi^{-}_I\\
\end{array}\right), ~
\phi_S=\frac{1}{\sqrt2}
\left(\begin{array}{c}
v_S+\rho_S+i\sigma_S\\
\sqrt2\phi^{-}_S\\
\end{array}\right),
\ee
we find the mass matrix 
of neutral Higgs bosons as
\be
H^{t}M^2_hH&=&
\frac{1}{2}\left(
\begin{array}{ccc}
\rho&\sigma&\varphi
\end{array}
\right)
\left(\begin{array}{ccc}
M^2_{\rho,\rho} & M^2_{\rho,\sigma} & 0\\
M^2_{\rho,\sigma}  &  M^2_{\sigma,\sigma} & 0\\
0  &  0 & M^2_{\varphi,\varphi}\\
\end{array}\right)
\left(\begin{array}{c}
\rho\\
\sigma \\
 \varphi\\
\end{array}\right),
\ee 
where $\rho=(\rho_I,~\rho_S)$, $\sigma=(\sigma_I,~\sigma_S)$.
Each $3\times 3$ element $M^2_{\rho,\rho}, M^2_{\rho,\sigma}, M^2_{\sigma,\sigma}$ are given by 
\cite{Kajiyama:2006ww}
\be
 M^2_{\rho,\rho}
&\simeq&
\left(\begin{array}{ccc}
0 & 2\mu^2_2 & \sqrt2{\rm Re}(\mu^2_4) \\
2\mu^2_2 & 0 & \sqrt2{\rm Re}(\mu^2_4) \\
\sqrt2{\rm Re}(\mu^2_4) & \sqrt2{\rm Re}(\mu^2_4) &0 \\
\end{array}\right)
+
\left( \begin{array}{ccc}
a_{\rho,\rho} v_D^2& a_{\rho,\rho} v_D^2 &b_{\rho,\rho}
v_D v_S \\ 
a_{\rho,\rho} v_D^2& a_{\rho,\rho} v_D^2&b_{\rho,\rho} v_D
v_S \\ 
b_{\rho,\rho} v_D v_S  &b_{\rho,\rho} v_D v_S
&c_{\rho,\rho} v_S^2\\
\end{array}\right)
,\\
 M^2_{\sigma,\sigma}
&\simeq&
\left(\begin{array}{ccc}
0 & 2\mu^2_2 & \sqrt2{\rm Re}(\mu^2_4) \\
2\mu^2_2 & 0 & \sqrt2{\rm Re}(\mu^2_4) \\
\sqrt2{\rm Re}(\mu^2_4) & \sqrt2{\rm Re}(\mu^2_4) &0 \\
\end{array}\right)\nonumber
\\&&\!\!\!\!\!
+\left( \begin{array}{ccc}
a_{\sigma,\sigma}v_D^2+a'_{\sigma,\sigma}v_S^2&
b_{\sigma,\sigma} v_D^2& c_{\sigma,\sigma}v_D v_S\\
b_{\sigma,\sigma}
v_D^2&a_{\sigma,\sigma}v_D^2+a'_{\sigma,\sigma}v_S^2&c_{\sigma,\sigma}v_D
v_S\\
c_{\sigma,\sigma}v_D v_S&c_{\sigma,\sigma}v_D
v_S&d_{\sigma, \sigma}v_D^2\\
\end{array}\right)
,\\
M^2_{\rho,\sigma}
&\simeq&
\left(\begin{array}{ccc}
0 & 0 & \sqrt2{\rm Im}(\mu^2_4) \\
0 & 0 & \sqrt2{\rm Im}(\mu^2_4) \\
\sqrt2{\rm Im}(\mu^2_4) & \sqrt2{\rm Im}(\mu^2_4) &0 \\
\end{array}\right)
+\left( \begin{array}{ccc}
a_{\rho,\sigma}v_S^2&0&-b_{\rho,\sigma}v_D v_S\\
0&a_{\rho,\sigma}v_S^2&-b_{\rho,\sigma}v_D v_S\\
b_{\rho,\sigma}v_D v_S&b_{\rho,\sigma}v_D v_S&c v_D^2\\
\end{array}\right), 
\ee
where the coefficients $a_{\rho, \rho}$'s are of ${\cal
O}(1)$.
The $\varphi$ term is given by
\bea
M^2_{\varphi,\varphi}
=
2m^2_2+v^2_S\lambda_2+v^2_D\lambda_3.
\eea 
Note that $\varphi$ is mass eigenstate automatically due to the exact
$Z_2$ symmetry.
The stable minimum conditions are found by partially differentiating the
potential by 
$\varphi$ as
\bea
\left.\frac{\partial V}{\partial \varphi}\right|_{\varphi\rightarrow 0}=0,~
\left.\frac{\partial^2 V}{\partial \varphi^2}\right|_{\varphi\rightarrow 
0}
=
M^2_{\varphi,\varphi},~
\left.\frac{\partial^2 V}{\partial \varphi\partial 
v_{S(D)}}\right|_{\varphi\rightarrow 0}=
\frac{1}{\sqrt{2}}v_{S(D)}m_{4(5)}.
\eea
Therefore, we simply obtain the vacuum conditions for $\langle 
\phi_{I,S}\rangle\neq 0$ and 
$\langle \varphi\rangle=0$ as $M^2_{\varphi,\varphi}>0$. Since $\varphi$ is mass eigenstate,
the mass matrix $M^2_h$ is diagonalized by the $7\times 7$ orthogonal 
matrix ${\cal O}$ which is decomposed into $6\times6$ and $1\times1$, as
${\cal O}M^2_h{\cal O}^T$. 
Notice that quarks couple only with $\phi_S$ via Yukawa interactions, 
and also that there is no mixing between 
$\phi$ and $\eta$ because $\eta_{I,S}$ do not get VEVs.

The SM Higgs is described in terms of the linear combination of flavor 
eigenstate fields as
\bea
{\mbox{SM-Higgs}}={\cal O}_{11}\rho_1+{\cal O}_{12}\rho_2+{\cal O}_{13}\rho_S+{\cal 
O}_{14}\sigma_1
+{\cal O}_{15}\sigma_2+{\cal O}_{16}\sigma_S, 
\eea
where we hereafter define the SM Higgs mass as $m_h$.  
The other combinations correspond to heavy neutral Higgs bosons with 
mass of ${\cal O}(1)$ TeV.
From Eq.(\ref{higgs-int}), we write down the key interacting term in the direct detection as
\bea
V(\phi,\varphi)\sim \lambda_2v_S{\cal O}_{31}\rho_S\varphi^2.
\eea

\section{Constraint from WMAP} 
\begin{figure}[htb]
\begin{center}
\includegraphics*[width=0.3\textwidth]{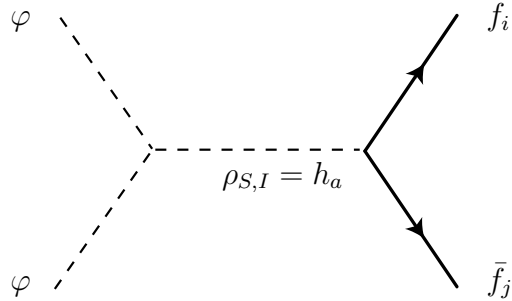}
\put(10,100){$f_i$}
\put(10,0){$\bar f_j$}
\put(-90,40){$\rho_{S,I}=h_a$}
\put(-170,100){$\varphi$}
\put(-170,0){$\varphi$}
\caption{\footnotesize
Annihilation diagrams of $\varphi$ for the cross section $\left<\sigma v\right>$. Where $f$ runs all the fermions; leptons and quarks whose masses are less than $M_\varphi$. 
}\label{diagrams}
\end{center}
\end{figure}

Our dark matter candidate $\varphi$ annihilates into fermion pair $f_i \bar f_j$, where $i,j$ are 
generation indices, via $h_a$-mediated s-channel diagram. 
There exist six Higgs bosons $h_a~(a=1-6)$ in our model. 
The relevant operators are originated from the Higgs potential and the Yukawa interactions, 
which are given by  
\be
{\cal L}&=&-\left[ \lambda_2 v_S {\cal O}_{a3}
+\sqrt{2}\lambda_3 v_D\left( {\cal O}_{a1}+{\cal O}_{a2}\right)\right]h_a \varphi^2\nonumber \\
&+&\frac{1}{\sqrt{2}}Y_{ui}\bar u_i\left[ \left( {\cal O}_{a3}-i{\cal O}_{a6}\right)P_L
+ \left( {\cal O}_{a3}+i{\cal O}_{a6}\right)P_R\right]u_i h_a\nonumber \\
&+&\frac{1}{\sqrt{2}}Y_{di}\bar d_i\left[ \left( {\cal O}_{a3}+i{\cal O}_{a6}\right)P_L
+ \left( {\cal O}_{a3}-i{\cal O}_{a6}\right)P_R\right]d_i h_a\nonumber \\
&+&\frac{1}{\sqrt{2}}\bar e_i \left[U_{eR}^{\dag} \left\{ (Y^{e1})^T\left( {\cal O}_{a1}+i {\cal O}_{a4}\right)+
(Y^{e2})^T\left( {\cal O}_{a2}+i {\cal O}_{a5}\right)\right\}U_{eL} \right]_{ij}P_Le_j h_a
\nonumber \\
&+&\frac{1}{\sqrt{2}}\bar e_i \left[U_{eL}^{\dag} \left\{ Y^{e1}\left( {\cal O}_{a1}-i {\cal O}_{a4}\right)+
Y^{e2}\left( {\cal O}_{a2}-i {\cal O}_{a5}\right)\right\}U_{eR} \right]_{ij}P_Re_j h_a\\
&\equiv& -\frac{1}{2}A_a h_a \varphi^2+\bar u_i \left( B_a P_L+B_a^*P_R\right)u_ih_a
+\bar d_i \left( C_a P_L+C_a^*P_R\right)d_ih_a\nonumber \\
&&+(\bar e_i D_{ij}^a P_L e_j h_a+h.c.),
\ee
where repeated indices are summed up as $a=1-6$, and $i,j=1-3$ for $M_{\varphi}>m_{i,j}$. 
We simply find the thermally averaged  cross section $\langle \sigma v 
\rangle$
for the annihilation of two $\varphi$'s \cite{griest1} from Fig.\ref{diagrams} 
\be
\langle \sigma v\rangle &=& a + b \frac{6}{x}+\cdots,~\\
a& =&\sum_{a,i,j}\sum_X \left| A_a\right|^2
 I_{2,i,j}
\frac{1}{4 M_{\varphi}^2 m_a^2 (m_a^2+\Gamma_a^2)}\nonumber\\
&\times& \left[ \left( \left| X_{ij}^a\right|^2+\left| X_{ji}^a\right|^2\right)\left( 4M_{\varphi}^2-m_i^2-m_j^2\right)
-2 m_i m_j \left( X_{ij}^a X_{ji}^a+h.c.\right)\right],\\
 ~b& =&-\frac{1}{4}a+\sum_{a,i,j}\sum_X \left| A_a\right|^2
 I_{2,i,j}
 \frac{1}{4 m_a^2 (m_a^2+\Gamma_a^2)}
 \left( \left| X_{ij}^a\right|^2+\left| X_{ji}^a\right|^2\right),
\label{bandr}\\
  I_{2,i,j}&=&\frac{1}{8\pi M^2_{\varphi}}\sqrt{\left(M^2_{\varphi}-(m_i+m_j)^2\right)
  \left(M^2_{\varphi}-(m_i-m_j)^2\right)},
  \label{randy}
\ee
where $M_\varphi$ is $\varphi$ mass which is our DM
candidate and the coupling $X_{ij}^a$ stands for $B_a\delta_{ij},~C_a \delta_{ij}$,
$D^a_{ij}$. 
The parameter $x$ is the ratio of the DM mass $M_{\varphi} $ and the temperature of the 
Universe $T$; $x=M_\varphi/T$. 
The mass parameters $m_i^2$ and $m_j^2$ are fermion masses of the final states, and 
$m_a^2 $ and $\Gamma_a$ are mass and decay width of the intermediating Higgs bosons, respectively. 

\begin{figure}[h]
 \begin{center}
\vspace{1cm}
\hspace{1.2cm}
\includegraphics[width=0.34\textwidth,angle=90]{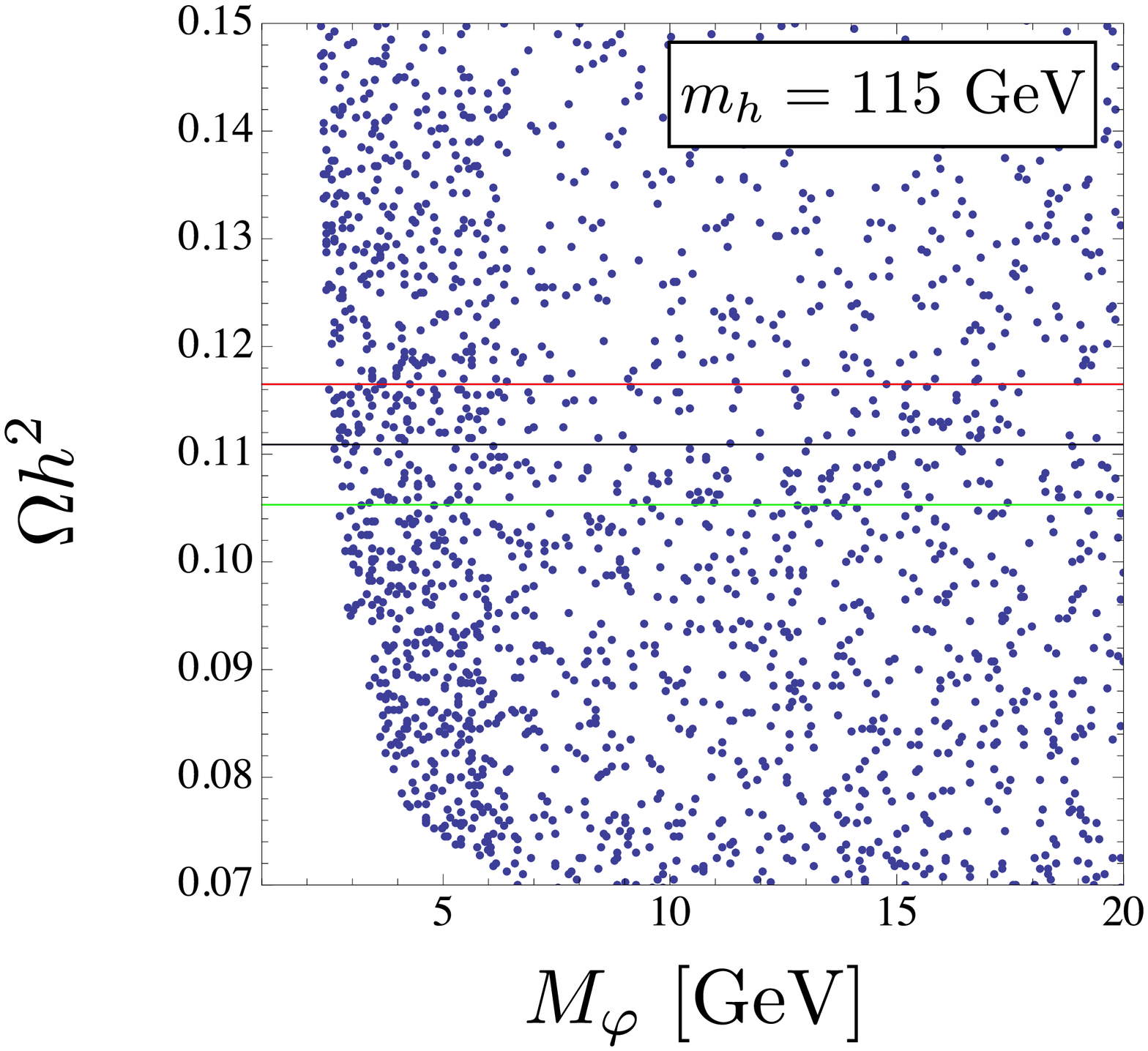}
\hspace{-0.8cm}
\includegraphics[width=0.34\textwidth,angle=90]{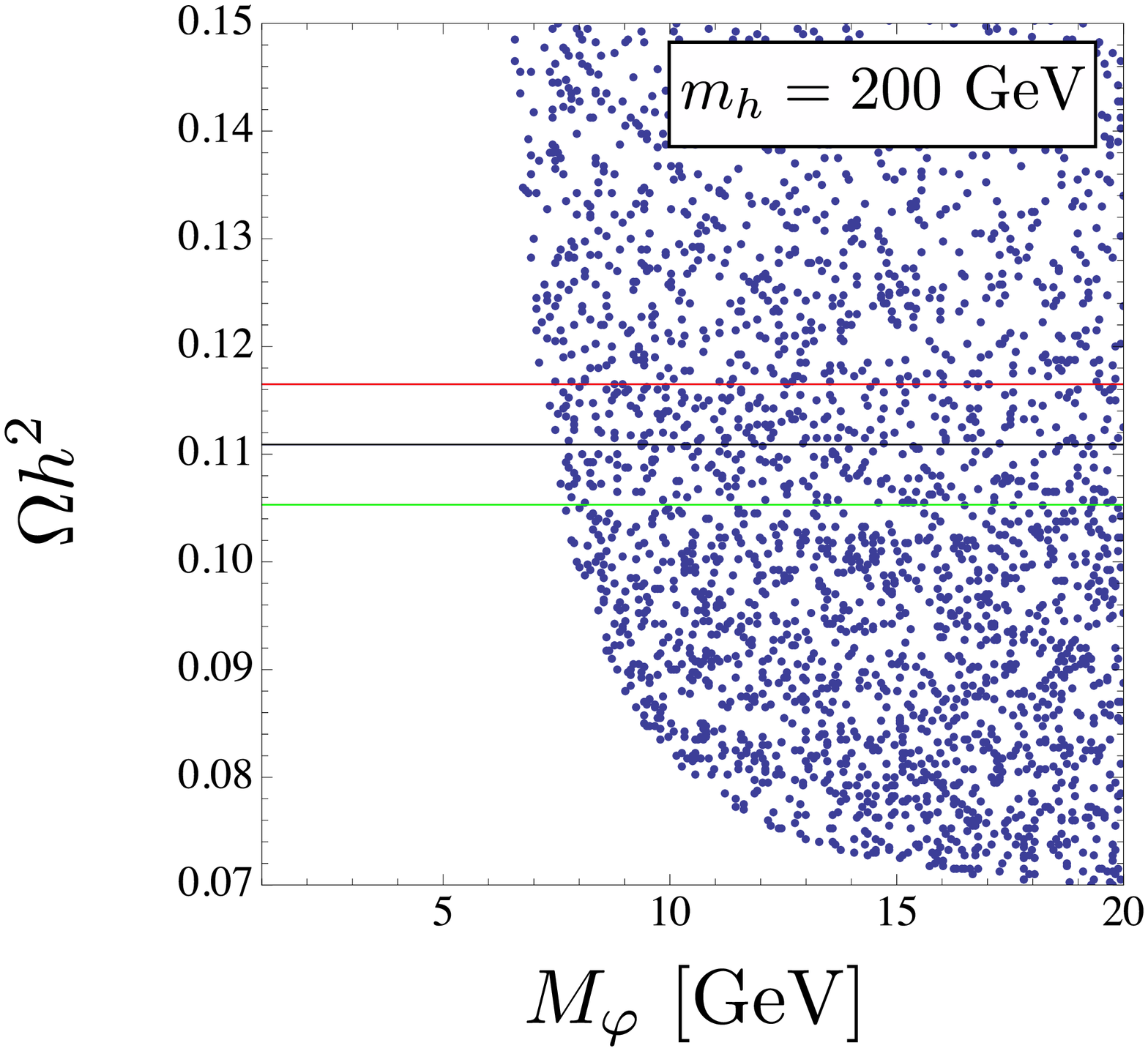}
\caption{\footnotesize
The allowed region in the $\Omega h^2-M_\varphi$ plane
in which $0.09\le\Omega_d h^2\le0.12$.
}\label{relic}
\end{center}
\end{figure}

In Fig. \ref{relic}, we show
the allowed region; $0.09\le\Omega_d h^2\le0.12$ at 3$\sigma$ \cite{Antoniadis:2010nb}, in the $\Omega h^2-M_\varphi$ plane, in which one finds that there is a wide allowed range. In our model, either of $|\lambda_{2,3}|$ is of ${\cal O}$(1) to find the promising points. Since we take $|\lambda_{2}|\simeq{\cal O}$(0.01) to fit the experiments from the direct detections in the next section,
$|\lambda_{3}|\simeq {\cal O}$(1) plays an important role for obtaining the relic abundance.
As can been seen from Fig.\ref{relic}, we find the allowed mass range as follows:
\be
&&
2\ {\rm GeV}<M_\varphi \quad {\rm for}\ {m_h=115}\ {\rm GeV},\\
&&
8\ {\rm GeV}<M_\varphi \quad {\rm for}\ {m_h=200}\ {\rm GeV}.
\ee
Notice in the figure that the range of $m_h$; $115-200$ GeV, is not forbidden by the current experiments of ATLAS and CMS
due to the mixing between multi-Higgs bosons, as can been shown in the section \ref{higgs}.

\section{Direct Detection}\label{secdirect}
\begin{figure}[h]
\begin{center}
\includegraphics[scale=1.0]{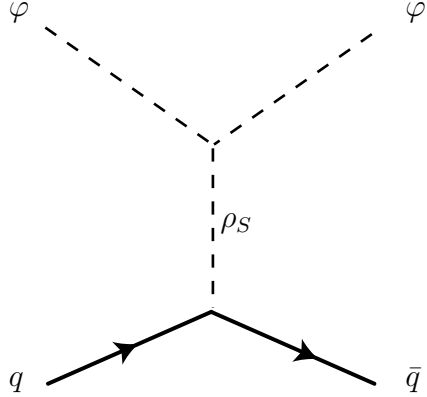}
\put(10,140){$\varphi$}
\put(10,0){$\bar q$}
\put(-60,60){$\rho_S$}
\put(-140,140){$\varphi$}
\put(-140,0){$q$}
\caption{The t-channel diagram for the direct detection of DM.}   
\label{dd-diag}
\end{center}
\end{figure}

We analyze the direct detection search of DM through the experiments of CoGeNT \cite{cogent}, DAMA \cite{Bernabei:2008yi}, including XENON100 
\cite{Aprile:2010um}. The global fit analysis of DM mass and elastic cross
section is done in Ref.\cite{Kopp:2009qt}. 
The main contribution to the spin-independent cross section comes from the 
t-channel diagram 
mediated by $\rho_S$, as depicted in 
Fig.\ref{dd-diag}. 
Then the resultant cross section for a proton is given by 
\be
\sigma^{(p)}_{SI}=\frac{4}{\pi}\left(\frac{m_p 
M_\varphi}{m_p+M_\varphi}\right)^2|f_p|^2,
\ee
with the hadronic matrix element
\be
\frac{f_p}{m_p}=\sum_{q=u,d,s}f^{(p)}_{T_q}\frac{\alpha_q}{m_q}
+\frac{2}{27}\sum_{q=c,b,t}f^{(p)}_{TG}\frac{\alpha_q}{m_q},
\ee
where $m_p$ is the proton mass \cite{Jungman:1995df, Belanger:2008sj}. 
The effective vertex $\alpha_q$ in our case 
is given by
\be
\alpha_q\simeq \frac{{\cal O}_{31}{\cal 
O}_{31}\lambda_2}{m^2_{h}}\frac{m_q}{M_\varphi}, 
\label{eq:alpha}
\ee
where $m_q$ is quark mass. Notice that since the quark sector couples only to $\phi_S$, 
the diagram mediated by the real part $\rho_S$ of $\phi_S$ gives dominant contribution.
\begin{figure}[t]
\begin{center}
\includegraphics[scale=0.70]{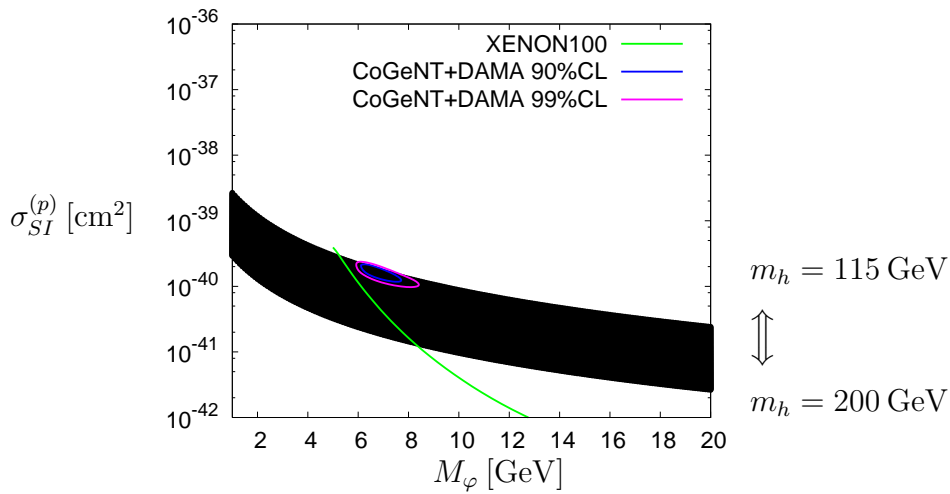}
\put(-280,90){${\tiny \sigma^{(p)}_{SI}\ [{\rm cm^2}]}$}
\put(-120,-7){${\tiny M_{\varphi}\ [{\rm GeV}]}$}
\put(0,70){${\tiny m_{h}=115\ {\rm GeV}}$}
\put(0,45){${\Big\Updownarrow}$}
\put(0,20){${\tiny m_{h}=200\ {\rm GeV}}$}
\caption{The spin-independent cross section as a function of the DM mass 
for the direct detection \cite{cogent,Bernabei:2008yi}. The longitudinal black line represents the SM Higgs boson mass range.}   
\label{cdms}
\end{center}
\end{figure}

In the numerical analysis, we set the Higgs masses to avoid the lepton 
flavor violation (LFV) process 
as follows:
\be
115\ {\rm GeV}\le m_{h}\le 200\ {\rm GeV},~1000\ {\rm GeV}\lesssim 
{\rm Other\ five\ 
neutral\ Higgs\ boson\ masses}.
\ee
We have investigated that we can choose the above parameter ranges by numerical calculation.
Under this setup, the elastic cross section is shown in Fig.\ref{cdms}. 
Where we set ${|\cal O}_{31}{\cal O}_{31}\lambda_2|=0.005$.
We plot the DM mass $M_\varphi$ in the region $1-20$ GeV. 
From Fig.\ref{cdms}, we can see that the lighter mass of SM Higgs 
is favored by CoGeNT and DAMA experiments in our parameter set. 


\section{Higgs Phenomenology}\label{higgs}
\begin{figure}[thb]
\begin{center}
\includegraphics[scale=0.50]{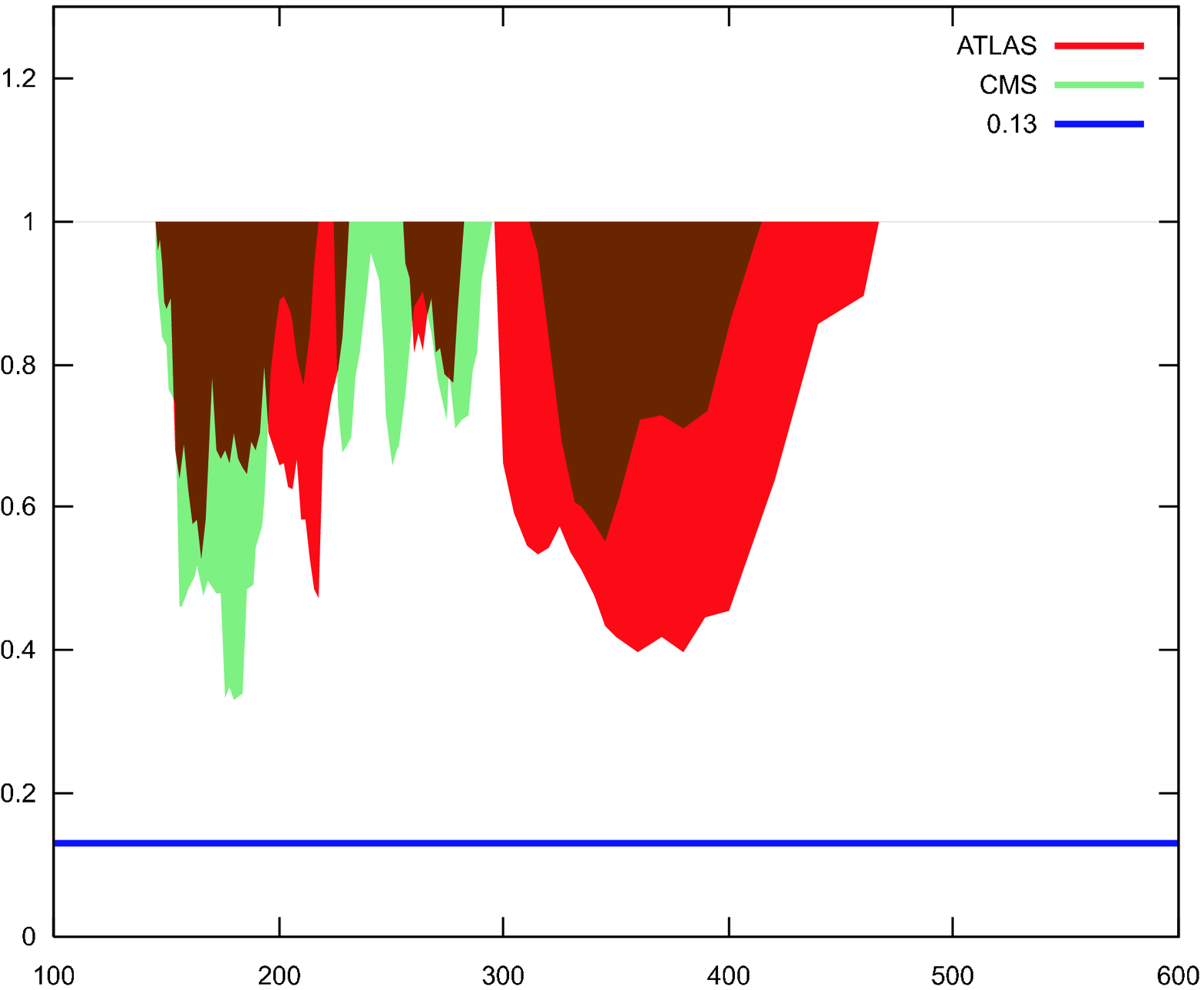}
\put(-320,110){${\tiny |{\cal O}_{31}|^4}$}
\put(-150,-13){${\tiny m_{h}\ {\rm GeV}}$}
\caption{The excluded regions of the ATLAS and CMS experiments: The
 red region is excluded by ATLAS, the light green region is excluded
 by CMS and the brown region is excluded by both experiments. 
The blue line of $|{\cal O}_{31}|^4=0.13$ is our benchmark point, which
 implies that our model is not constrained by both experiments.}
\label{cms-atlas}
\end{center}
\end{figure}

Recently ATLAS and CMS reported the upper limit of the SM Higgs mass, in which there are no significant excess
in the range around $145\le m_h\le 466$ GeV. However once there are mixing between SM Higgs and the other neutral bosons,
such an upper bound could be relaxed due to the coefficient of the mixing. In our case, actually, since we have the ${\cal O}_{31}$
coefficient, the constraint is dramatically relaxed \footnote{We would like to thank Jong-Chul Park for the useful discussions}. In Fig. \ref{cms-atlas}, one finds that there is no constraint from ATLAS and CMS. Where we take $|{\cal O}_{31}|^4\simeq 0.13$ in the direct detection benchmark.  

\begin{figure}[thb]
\begin{center}
\includegraphics[scale=0.58]{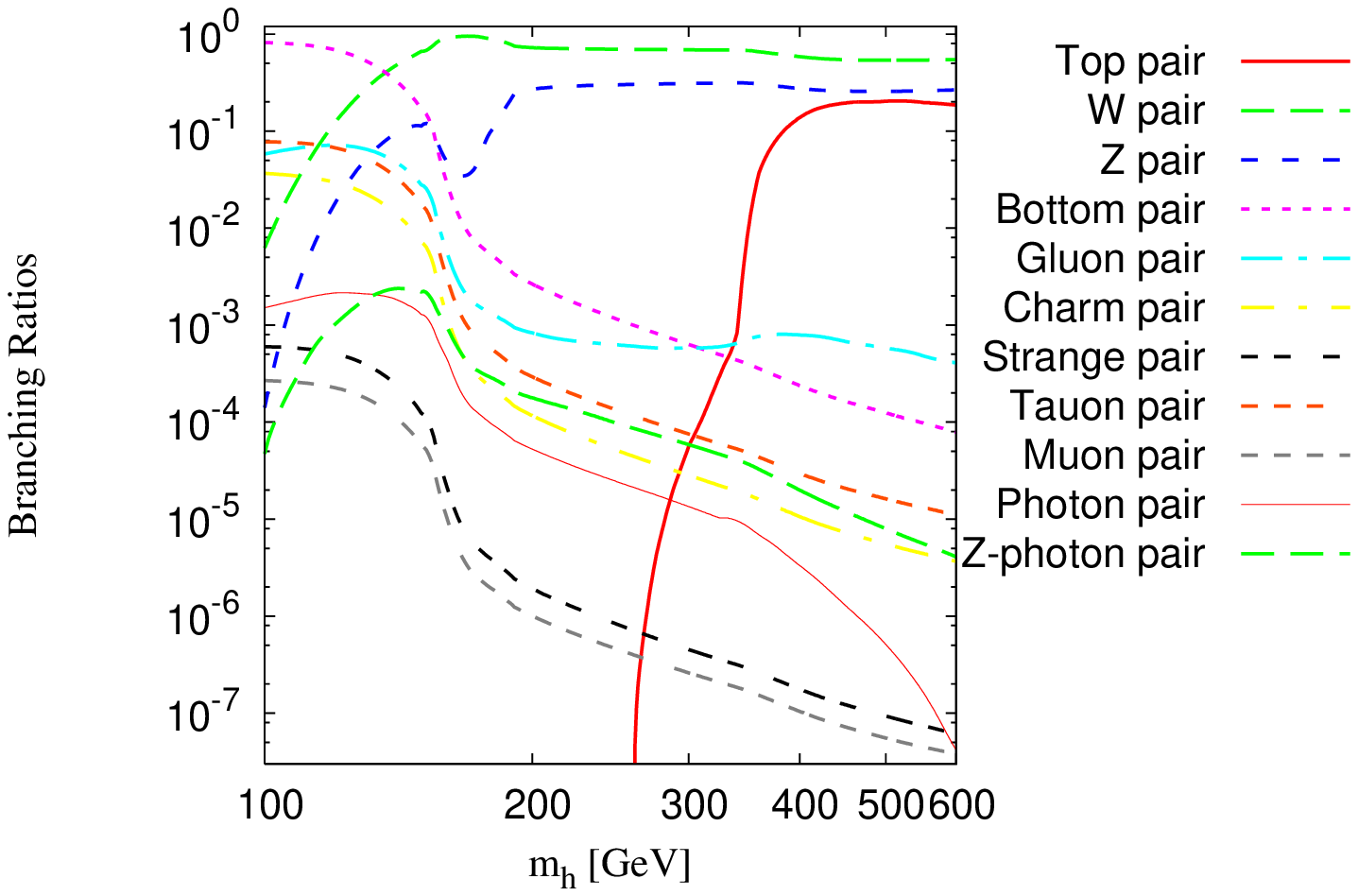}
\includegraphics[scale=0.58]{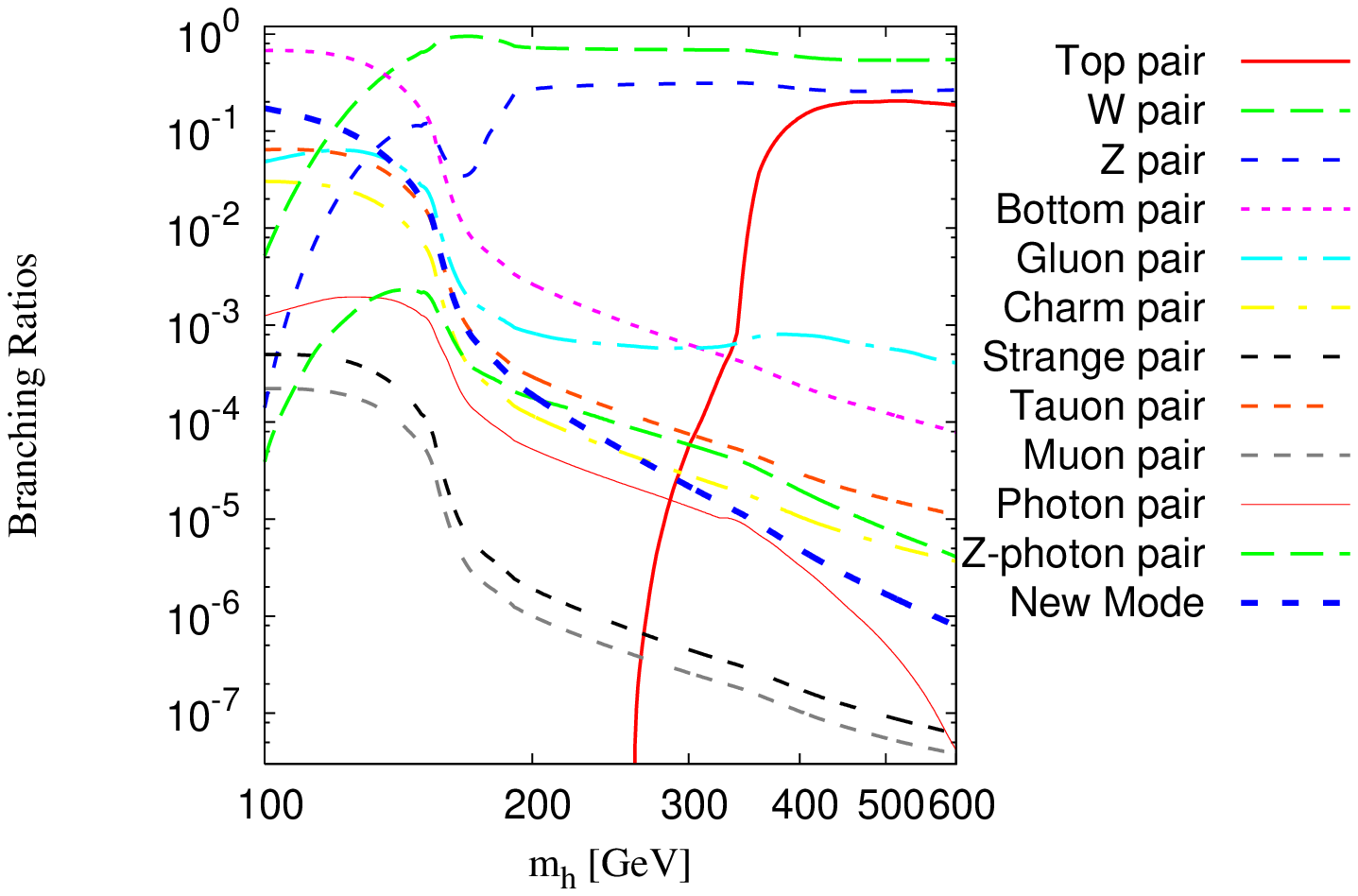}
\caption{The left panel is depicted the branching ratio of the SM Higgs boson.
The right panel is depicted the branching ratio of the our SM Higgs boson: The new contribution (blue thick dashed line) is dominant for the lower range of $m_h$,
then it goes down for the higher range.}   \label{brhiggs}
\end{center}
\end{figure}

In case of decay, 
our SM Higgs \cite {Djouadi:2005gi} has a new channel of $h\rightarrow \varphi\varphi$ whose vertex is proportional to $v_s {\cal O}_{31}\lambda_2$.
One finds that it affects on the branching ratios of the Higgs by comparing the left and right panel in Fig. \ref{brhiggs}.
In particular, the new contribution could be second dominant for the lower range of $m_h$,
then it goes down for the higher range, as can been seen from the right panel of Fig. \ref{brhiggs}.

\section{Conclusions}
We have considered the rather light DM in favor of the direct detection recently reported by CoGeNT, DAMA (and XENON100)
in a $D_6$ symmetric radiative seesaw model. 
We found that a gauge and $D_6$ singlet scalar boson $\varphi$ can be a promising DM candidate in the ragion $1-20$ GeV
and be consistent with the WMAP experiment. Together with them, one finds that
rather smaller SM Higgs mass is favored if these experiments could detect the DM near the current bound.
We have also shown that our SM Higgs mass bound recently reported by the
ATLAS and CMS experiments can be escaped due to the mixing between SM
Higgs and other neutral bosons. In our benchmark of the direct
detection, especially, we found that 
the Higgs mass is not constrained throught the both of the experiments.

\section*{Acknowledgments}
H.O. thanks to Prof. Eung-Jin Chun, Dr. Jong-Chul Park, and Dr. Priyotosh Bandyopadhyay
for fruitful discussion of Higgs phenomenologies.
This work is supported by 
Young Researcher Overseas Visits Program for Vitalizing Brain Circulation 
Japanese in JSPS (T.T.). 
H.O.\ acknowledges
partial supports from the Science and Technology Development Fund
(STDF) project ID 437 and the ICTP project ID 30.


\end{document}